\documentclass[10pt]{blois07}
\usepackage{graphicx}
\usepackage{amssymb} 
\usepackage{amsmath}
\usepackage{cite,./mcite}

\newcommand{\jet}{{\cal J}}
\newcommand{\pdf}{{\cal F}}
\newcommand{\soft}{{\cal S}}
\newcommand{\hardfact}{\ensuremath{H}}
\begin{document}
\title{Blois07/EDS07 Proceedings \\ Unintegrated parton distributions and correlation functions}
\author{Anna Sta\'sto}
\institute{Penn State University, 104 Davey Lab., University Park, 16802 PA, USA}
\maketitle

\begin{abstract}
We discuss the limitations of the standard collinear approach. The kinematical approximations
necessary to derive the collinear factorization are insufficient for the description of the exclusive final states.
We argue that for a proper treatment of the final states one needs to use fully unintegrated parton correlation functions.
We  introduce the gauge invariant definitions of these objects and the factorization theorem for one jet production
in deep inelastic scattering.
\end{abstract}

%%%%%%%%%%%%%%%%%%%%%%%%%%%%%%%%%%%%%%%%
\section{Introduction: factorization and integrated parton distribution functions}
\label{sec:intro}

The leading twist formalism based on the collinear factorization \cite{Collins:1989gx}  has been proved widely successful in 
a variety of processes in QCD. The collinear factorization allows for the separation of the short and long distance contributions in the processes which involve hadrons. The short distance part is the partonic cross section, which can be systematically 
calculated order by order in the strong coupling  within the  perturbation theory. The long distance parts are parton distribution and fragmentation functions which contain all the non-perturbative information. Thanks to the factorization theorem these quantities are universal and once determined in one process can be taken over to the other process to determine the corresponding cross section.   These standard parton distribution functions are often referred as the integrated parton distribution functions (PDFs) since they depend
only on the fraction of the longitudinal momentum $x$ and the hard scale $\mu$. 
Problems arise when one tries to address more exclusive processes with final states. Then, the details of the kinematics are essential and standard kinematic approximation can be insufficient  for these purposes.  This is where  the formalism with the parton distributions unintegrated over the other components of the momenta is necessary.  One also needs to prove more general factorization theorems which incorporate these unintegrated parton distribution/fragmentation functions.  

To start with, let us first recall the standard definition  of the integrated quark parton density:
\begin{equation}
f(x_{Bj},\mu) \; = \; \int \frac{dy^-}{4\pi} e^{-i x_{Bj}p^+y^-} \langle P | \bar{\psi}(0,y^-,0_T) V_y^{\dagger}(n) V_0(n) \gamma^+ \psi(0) |P \rangle_R \; .
\end{equation}
Here, $\psi$ is the quark field and  $| P \rangle $ is the proton state in which the  operator above is evaluated. Subscript $R$ means that we are considering renormalized operator and $\mu$ is the renormalization scale. The object inserted  between the quark fields is the Wilson line 
\begin{equation}
V_y^{\dagger}(n)V_0(n) \; = \; {\rm P} \exp \left (-ig \int_0^{y^-} d\lambda \, n \cdot A(\lambda n) \right) \; .
\end{equation}
 in the lightlike direction $n=(0,1,0_T)$. This ensures the gauge invariance and can be shown to arise from graphs with arbitrarily many gluon exchanges which appropriately factorize into this path-ordered exponential. 
 
\begin{figure}[h]
\centerline{\includegraphics[width=4cm]{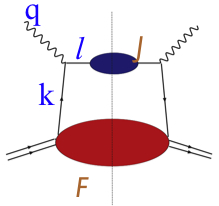}}
\label{fig:pm}
\caption{Parton model in deep inelastic scattering.}
\end{figure}

 To illustrate some issues related with the kinematical approximations used in the  standard collinear  approach let us consider the parton model of the deep inelastic scattering. The lowest order graph is pictured in Fig.~\ref{fig:pm} and we can write down formally its contribution  to the hadronic tensor as 
 \begin{equation}
 W^{\mu\nu}(q,P) \; = \;  \sum_j \frac{e^2_j}{4\pi} \int \frac{d^4k}{(2\pi)^4} {\rm Tr}[\gamma^{\mu} J_j(k+q)\gamma^{\nu}F_j(k,P)] \; ,
 \label{eq:hadtensor}
 \end{equation} 
with the sum over $j$ as a sum over the quark flavors. 

The kinematic approximations needed to obtain the factorized form
set the quarks on-shell and massless
\begin{eqnarray}
k\simeq \hat{k} &  \equiv &  (x_{Bj}P^+,0,0_T) \; , \nonumber \\
l \simeq \hat{l} & \equiv  &(0,\frac{Q^2}{2x_{Bj}P^+},0_T) \; .
 \label{kin:approx}
\end{eqnarray}
These approximations give the  familiar expression for the hadronic tensor 
\begin{equation}
W^{\mu\nu}(q,P) \; \simeq \;  \sum_j \frac{e^2_j}{4\pi}  \bigg \{ \int  \frac{dk^- d^2k_T}{(2\pi)^4} F_j^+(x_{Bj}P^+,k^-,k_T) \bigg \} \; {\rm Tr}[\gamma^{\mu}\gamma^+\gamma^{\nu}\hat{k}\!\!\!/] \; .
\label{eq:hadtensorfacrorized}
\end{equation}
The expression in the curly brackets corresponds to the parton density and the expression with the trace is the partonic cross section. Note, that the approximations are good when the momentum components $k_T$ and $k^-$ are small with respect to $k^+$.  This is sufficient when the momenta are integrated over in the inclusive cross section but might lead to large errors when we look at some more exclusive processes.  Indeed the parton model approximation makes the following replacement
$$
\delta^{(4)}(k+q-l) \longrightarrow \delta(k^+ + q^+)\delta(q^--l^-)\delta^{(2)}(l_T) \; .
$$
These two delta functions can lead to  similar results when integrated over all momenta (inclusive processes) but 
can lead to very different results in the exclusive cases. If we allow for the invariant mass  of the outgoing parton $l^2=M^2$ we see that the parton model value $k^+=x P^+$ becomes significantly modified
$$
k^+=x P^++\frac{M^2+k_T^2}{2(k^-+q^-)} \; .
$$
As shown in \cite{Collins:2005uv} on the example of the $c\bar{c}$ production in DIS there are 
huge differences between standard parton model, unintegrated partons and exact kinematics.
In particular the $p_T$ distributions of the produced charmed quarks depend heavily whether integrated or unintegrated formalism is used.
The general conclusion  \cite{Collins:2005uv} is that conventional formalism with integrated parton densities is not suitable for the analysis of the final states. In the next sections we discuss the general factorization framework which makes use of the unintegrated parton correlation functions. All  the details, including the derivation of the factorization and the definitions of the unintegrated parton correlation functions can be found in \cite{Collins:2007ph}.

%%%%%%%%%%%%%%%%%%%%%%%%%%%%%%%%
\section{Integrated, unintegrated, fully unintegrated parton correlation functions}

We have learned from the discussion in the introduction that the description of the exclusive processes requires
a new formalism with new parton densities and fragmentation functions. We distinguish three types of objects:
\begin{itemize}
\item Conventional integrated parton distributions and fragmentation functions. These depend on the fraction of the  longitudinal momentum and the scale $$ {\cal F}(x,\mu) $$
\item Unintegrated parton distribution functions (and FFs) which depend on the transverse momentum
$${\cal F}(x,k_t,\mu)  $$
\item Fully unintegrated parton correlation functions which depend on all components of the momenta: longitudinal, transverse, and virtuality
$${\cal F}(x,k_T,k^2,\mu) $$
\end{itemize}

The unintegrated (over $k_T$) parton distribution functions have been introduced in the context of small $x$
physics, (see for example \cite{Lipatov:1996ts}).
There, as a result of the high-energy approximation, the 4-point gluon Green's function emerges as a fundamental object. It depends on the transverse momenta of 4-off shell gluons. Its evolution with rapidity is then governed by the BFKL equation.  There have been numerous efforts \cite{Kimber:2000bg,Kimber:2001sc,Watt:2003mx} to construct the unintegrated parton densities outside the small $x$ limit, many of them using the angular ordered CCFM equation.  However,  theses approaches do not provide with the operator definitions of the unintegrated parton distribution functions. Therefore, it is desirable to have  a unique and consistent approach which demonstrates factorization using the properly defined unintegrated parton correlation functions. Then, hopefully, different approaches developed so far will be reproduced as  a particular limit or approximation to a more general case.

%%%%%%%%%%%%%%%%%%%%%%%%%%%%%%%%%
\section{General strategy}
The general formalism for use in the case of the deep inelastic scattering  introduced in detail in \cite{Collins:2007ph}. The main points are
\begin{itemize}
\item Retain the exact kinematics in initial and final states.
\item Keep the explicit factors (bubbles) representing final states.
\item Retain on-shell matrix elements.
\item Define projections from exact to approximate momenta.
\item Construct definitions of the gauge invariant parton correlation factors  and the soft factor.
\item Use eikonal lines and Ward identities to prove factorization.
\item Use non-light-like eikonal Wilson lines to regularize light-cone divergences in the unintegrated parton distributions ( this introduces a cutoff in rapidity, similarly to what is done at small x ).
\end{itemize}

%%%%%%%%%%%%%%%%%%%%%%%%%%%%%%%%%
\begin{figure}[t]
\centerline{\includegraphics[width=7cm]{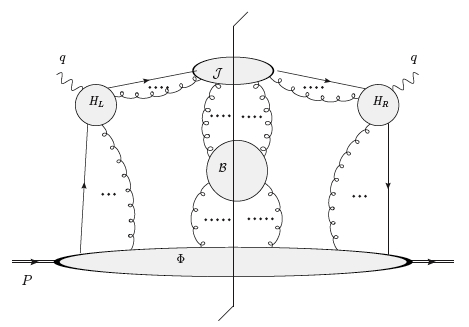}}
\label{fig:gengraph}
\caption{Most general graph with one jet production in deep inelastic scattering.}
\end{figure}
%%%%%%%%%%%%%%%%%%%%%%%%%%%%%%%%%
One starts with the most general graph in the case of the deep inelastic scattering which is illustrated in Fig.~\ref{fig:gengraph}. This graphs differs from the ordinary parton model due to several distinctive features. First of all, the final state quark has a jet subgraph. This is a minimal required assumption since we know that the quark has to hadronize.
A second important point is that we need to allow for the non-perturbative (or soft) interactions between the outgoing jet and the target remnant. These are obviously necessary to neutralize the color. Therefore we need to allow for arbitrarily many gluon connections between the soft bubble  and the target and jet subgraphs.  So graph in Fig.~\ref{fig:gengraph} constitutes  a necessary extension to a simple parton model.

%%%%%%%%%%%%%%%%%%%%%%%%%%%%%%%%%
\begin{figure}[t]
\centerline{\includegraphics[width=8cm]{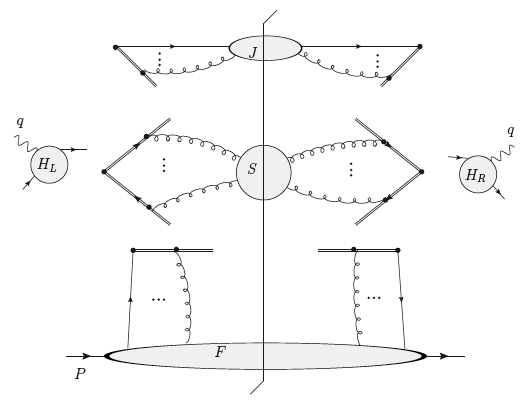}}
\label{fig:gengraphfact}
\caption{Most general graph with one jet production in deep inelastic scattering in the factorized form. Double lines denote the eikonal lines.}
\end{figure}
%%%%%%%%%%%%%%%%%%%%%%%%%%%%%%%%%
In order to prove  the factorization one needs first to identify all the leading regions: collinear to jet, collinear to the target and soft. On also needs to perform subtractions to ensure that the smaller regions are appropriately suppressed.
Then the Ward identities are used to disentangle the gluons between the different subgraphs (for example collinear to soft subgraphs). In particular one has to prove that the Ward identities work properly in the presence of the subtractions.
After the application of Ward identities the gluons factorize  into the Wilson lines in the appropriate objects.
The final factorization formula proven in \cite{Collins:2007ph} reads
\begin{multline}
  P_{\mu \nu} W^{\mu\nu}
  = 
  \int \frac{ d^4k_{\rm T} }{ (2\pi)^4 }
    \frac{ d^4k_{\rm J} }{ (2\pi)^4 }
    \frac{ d^4k_S }{ (2\pi)^4 }
    (2\pi)^4 \delta^{(4)}(q+P-k_{\rm T}-k_{\rm J}-k_S)
\times \\ \times 
    |\hardfact(Q,\mu)|^2
    ~ \soft(k_S,y_{\rm T},y_{\rm J},\mu) 
    ~ \pdf(k_{\rm T},y_p,y_{\rm T},y_s,\mu)
    ~ \jet(k_{\rm J},y_{\rm J},y_s,\mu),
\end{multline}
where $\soft,\pdf,\jet,\hardfact$ denote the soft factor, parton distribution, jet factor and hard scattering cross section respectively. $P^{\mu\nu}$ denotes the projection onto the apropriate structure function. All the non-perturbative factors depend on the all components of the 4-momenta $k_i$.  Note the appearance of the rapidity variables $y_i$.
These are necessary to  suppress the contributions from the regions where the rapidities are very large (rapidity divergencies). It has been demonstrated that this can be regularized by changing the direction of the Wilson lines from the light-like directions to slightly non-light-like   \cite{Collins:2007ph}.  The parton distribution, fragmentation functions and soft factors
acquire then additional parameters. Appropriate evolution equations give the prescription on the variation of these quantities with the rapidity.
The final factorized graph is shown in Fig.~\ref{fig:gengraphfact} where by double lines we denote the Wilson lines.

%%%%%%%%%%%%%%%%%%%%%%%%%%%%%%%%%
\section{Gauge invariant definitions of PCFs}
The analysis of the single gluon attachment allows to construct the gauge invariant definitions of the parton correlation functions (PCFs) and the soft factor.  As already discussed in a previous section critical to these are the directions of the Wilson lines. Soft gluons couple to the target jet, with 
its large plus component of
momentum, and to the outgoing jet with 
its large minus component of
momentum.  This means  that in coordinate space the soft factor is the vacuum
expectation value 
of Wilson lines that are nearly light-like in the plus
and minus directions \cite{Collins:2007ph}

\begin{multline}
  \tilde{\soft}(w,y_{\rm T},y_{\rm J},\mu) 
=
  \langle 0| I^{\dagger}_{n_{\rm T};w,0}  
   V_{w}(n_{\rm T}) V^{\dagger}_{w}(n_{\rm J}) I_{n_{\rm J};w,0}V_{0}(n_{\rm J}) 
   V^{\dagger}_{0}(n_{\rm T})  | 0 \rangle_{R} \; .
\label{softdef}
\end{multline}
 Note that we have replaced in this expression the light-like Wilson lines with the non-light-like ones in the directions
 denoted by the vectors $n_T=(1,-e^{-2y_{\rm T}},{\bf 0}_{T}),$ and $n_J=(-e^{-|2y_{\rm J}|},1,{\bf 0}_{T})$
 where $y_T \gg 1$ and $|y_J| \gg 1,\; y_J <0 $.
 As mentioned above, the reason for their presence is that they provide  cutoffs on rapidity
divergences. The factors $I$ are the transverse links at infinity which are needed to ensure strict gauge invariance
\cite{Belitsky:2002sm}.  With these included,  the soft factor (\ref{softdef}) is the expectation value of the closed Wilson loop.

The target PCF is defined as the gauge invariant expectation value  of two quark fields. Its definition reads \cite{Collins:2007ph}
\begin{multline}
\tilde{F}(w,y_p,y_{\rm T},y_s,\mu) 
=  
\frac{ \langle p | \bar{\psi}(w) V^{\dagger}_{w}(n_s)  I_{n_s;w,0}  
           \dfrac{\gamma^{+}}{2}
           V_{0}(n_s) \psi(0) | p \rangle_{R}
     }
     { \langle 0| I^{\dagger}_{n_{\rm T};w,0}  
            V_{w}(n_{\rm T}) V^{\dagger}_{w}(n_s) I_{n_s;w,0}
            V_{0}(n_s) V^{\dagger}_{0}(n_{\rm T})  | 0 \rangle_{R} 
     }.  
\label{pcf2}
\end{multline}
where $w$ is a reference point in space-time. The two quark fields are connected via three Wilson lines, two of them going into the direction $n_s  = (-e^{y_s}, e^{-y_s}, {\bf 0}_T)$ which corresponds to the rapidity close to zero in the center-of-mass system.
This expectation value is divided by the soft factor, in order to ensure the proper cancellation of the double counting contributions.
Note that the soft factor has now $n_J$ replaced by the $n_s$ vector. This guarantees that the definition (\ref{pcf2})
gives a good approximation for the gluons close to the target region.
Similar definition can be constructed for the jet factor with  vector $n_T$ replaced by $n_J$

\begin{equation}
\tilde{\jet}(w,\mbox{argument like $y_J$},y_{\rm J},y_s,\mu)
 = \frac{ \langle 0 | \bar{\psi}(w) V^{\dagger}_{w}(-n_s)  
          I_{-n_s;w,0}  \gamma^{-} V_{0}(-n_s) \psi(0)
          | 0 \rangle_{R}
        }
        { \langle 0| I^{\dagger}_{-n_s;w,0}  V_{w}(-n_s) V^{\dagger}_{w}(n_{\rm J}) I_{n_{\rm J};w,0}V_{0}(n_{\rm J}) 
               V^{\dagger}_{0}(-n_s)  | 0 \rangle_{R} 
        } .
\label{pcf3}
\end{equation}

%%%%%%%%%%%%%%%%%%%%%%%%%%%%%%%%%
\section*{Acknowledgments}
The work presented in this talk was performed in collaboration with John Collins and Ted Rogers \cite{Collins:2007ph}.
I would like to thank the organizers of EDS2007 for the kind invitation to this  very interesting workshop.
This research was supported by the 
U.S. D.O.E. under grant number DE-FG02-90ER-40577 and by the Polish Committee for Scientific Research under  grant No.\ KBN 1 P03B 028 28.

%------------------------------------------------------------------------------
%       Bibliography
%------------------------------------------------------------------------------
\begin{footnotesize}
\bibliographystyle{blois07} 
{\raggedright
\bibliography{blois07}
}
\end{footnotesize}
\end{document}